\documentclass[aps,pre,superscriptaddress,twocolumn,balancelastpage]{revtex4-1}

\usepackage[colorlinks,bookmarks=false,citecolor=blue,linkcolor=blue,urlcolor=blue]{hyperref}
\usepackage[all]{hypcap}   

\usepackage{amsmath,amssymb}
\usepackage{graphicx}
\graphicspath{{figures/}{/}}

\usepackage{verbatim}
\usepackage{color}
\usepackage{tikz}
\usepackage{lipsum}
\usepackage{dsfont}
\usepackage{placeins}  
\usepackage{flafter}
\usepackage{physics}
\usepackage{color}

\usepackage{enumitem}

\newcommand{\ue}{\text{e}}
\newcommand{\ui}{\text{i}}

\newcommand{\ud}{\text{d}}

\newcommand{\Th}{T_\text{H}}
\newcommand{\undga}{\underline{\gamma}}

\makeatletter
\let\Hy@backout\@gobble
\makeatother

\begin{document}

\title{Semiclassical foundation of universality in chaotic quantum circuits}

\author{Maximilian F.~I. Kieler}
\affiliation{CESAM research unit, University of Liège, B-4000 Liège, Belgium}
\affiliation{TU Dresden, Institute of Theoretical Physics and Center for
Dynamics, 01062 Dresden, Germany}

\author{Felix Fritzsch}
\affiliation{Max Planck Institute for the Physics of Complex Systems, 01187
Dresden, Germany}

\author{Arnd B\"acker}
\affiliation{TU Dresden, Institute of Theoretical Physics and Center for
Dynamics, 01062 Dresden, Germany}

\date{\today}
\pacs{}

\begin{abstract}
The fundamental correspondence between quantum chaotic single-particle systems
and random matrix theory is well-understood via periodic orbit theory.
In contrast, we show that many-body systems with explicit subsystem
structure possess characteristics different from the single-particle theory.
We present a periodic orbit theory for many-body systems with well defined
semiclassical limit.
For this we identify periodic orbit families arising exclusively in the
many-body setting and implement a central limit theorem characterizing their
correlations.
Based on this we demonstrate that spectral correlations in chaotic quantum
circuits
are characterized by the breaking of individual time translation invariance of
periodic orbits in the subsystems into residual synchronous time translations
only.
This provides a systematic approach to confirming random matrix universality
in deterministic many-body systems.
\end{abstract}

\maketitle

The detailed understanding of emergent random matrix behavior in quantum
many-body systems is crucial for establishing a quantum chaotic phase in
condensed matter physics. Spectral statistics and in particular the spectral
form factor (SFF) are important quantities to detect and quantify quantum
chaos in the many-body setting \cite{KosLjuPro2018, SunBonProVid2020,
ChaDeCha2021, BerKosPro2022}. This covers both systems with and without a
semiclassical limit ranging from many-particle system \cite{RoyPro2020,
RoyMisPro2022, IkeVidFly2025}, dual-unitary quantum circuits
\cite{AkiWalGutGuh2016, BerKosPro2018, FlaBerPro2020, BerKosPro2021}, quantum
computers \cite{DonZhaEtAl2025, VasGraBarSieZol2020} to quantum gravity
\cite{SaaSheSta2019:p, SaaStaYanYao2024, HanUrbMorWebRic2025}.
The motivation for studying quantum chaos based on spectral statistics is
founded upon the Bohigas-Giannoni-Schmit (BGS) conjecture \cite{BohGiaSch1984,
BohGiaSch1984b}, which postulates a fundamental relation between systems with
a chaotic semiclassical limit and random matrix theory (RMT).
The semiclassical foundations of this conjecture are well understood through
the seminal research over several decades collected under the name
\textit{periodic orbit theory} \cite{HaaGnuKus2018, Gut1971, Tab1983a,
HanAlm1984, Ber1985, SieRic2001, MueHeuBraHaaAlt2004, MueHeuBraHaaAlt2005,
MueHeuAltBraHaa2009}.
It provides a detailed picture of emergent RMT behavior in single-particle
systems with a semiclassical limit by connecting the SFF to correlations of
classical periodic orbits. In particular, it is shown that RMT behavior arises
as consequence of the time-translation symmetry of periodic orbits
\cite{Gut1971, Tab1983a, HanAlm1984, Ber1985} as well as higher order orbit
correlations originating from self-encounter bunches \cite{SieRic2001,
MueHeuBraHaaAlt2004, MueHeuBraHaaAlt2005, MueHeuAltBraHaa2009}.
This picture can be lifted to genuine many-body systems with a quasi-classical
limit, e.g.,~Bose Hubbard systems \cite{EngUrbRic2015, DubMue2016,
RicUrbTom2022}. More generally, periodic orbit theory is well suited for
describing systems with  multiple effectively \textit{all-to-all interacting}
degrees of freedom. In contrast, its application to many-body body systems with
\textit{an explicit subsystem structure}, which is ubiquitous in, e.g., condensed
matter settings, even though qualitatively conjectured
\cite{LiaGal2022}, is widely unexplored.

\begin{figure}
\includegraphics[]{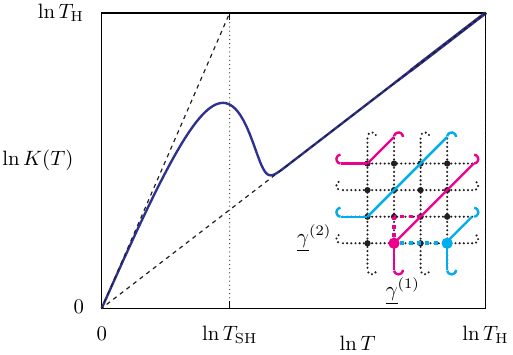}
 \caption{Bump-ramp structure of the SFF, see \eqref{eq:sff_fin_res_unscaled}
for $\chi = 0.975, L=3$ (blue, full) in double-logarithmic scale.
The black dashed lines indicate the limiting cases $K(T) = T^L$ (for $\chi =
1$) and $K(T) = T$ (for $\chi = 0$). Inset: Schematic structure of a family
$\Gamma_0 = (\undga^{(1)}, \undga^{(2)})$ (magenta dot) for $L=2$ and $T=4$.
The magenta and cyan curves indicate two periodic orbits living on $\Gamma_0$,
generated by $\phi_0^{(t, t)} \Gamma_0$ and $\phi_0^{(t, t+2)} \Gamma_0$,
respectively. The synchronous time translation $\phi_0^{(1, 1)}$ is indicated
by a magenta dashed line and an asynchronous time translation $\phi^{(0, 2)}$
is indicated by a cyan dashed line.
}
 \label{fig:periodic_orbits}
\end{figure}

In this paper, we present explicitly the process by which universal RMT
behavior emerges in interacting many-body quantum maps with a semiclassical
limit. Conceptually, this process can be understood as a collective
synchronization of orbit correlations in the subsystems towards synchronized
orbit correlations of the full system \cite{LiaGal2022, AltTelMic2024,
AltKimMic2025:p}. More precisely, the local time-translation symmetries of the
periodic orbits in the individual subsystems, giving rise to a non-trivial
short time regime of the SFF (see Fig.~\ref{fig:periodic_orbits}), break down to a global
time-translation symmetry only, reproducing the full RMT result.
We emphasize that we give here a concrete formulation for deterministic systems
and thereby complement recent investigations of corresponding random matrix
models based on the Weingarten calculus \cite{ChaLucCha2018a,
ChaLucCha2018b, FriChaDeCha2019, ChaDeCha2021,GarCha2021,ChaShiHusDe2022,
WuShiCha2025:p, YosGarCha2025, FriKie2024} and field theoretic methods
\cite{WinSwi2022, LiaGal2022, AltTelMic2024, AltKimMic2025:p}. Thus we
explicitly establish the intimate correspondence of quantum chaos and random
matrix theory even in the realm of spatially structured many-body systems.

\paragraph*{Setting}
We consider $L$-partite chaotic quantum maps \cite{BerBalTabVor1979}, which
can be seen as the deterministic pendants to the random matrix circuit models.
Given the position basis of the many-body Hilbert space $\mathcal{H}^{\otimes
L}$ as $\{ \ket{q} \}$ with $q = (q^{(1)}, \ldots, q^{(L)})$, we define the
one-step time-evolution operator of such a quantum map by
\begin{align}
    \matrixel{q'}{U(\epsilon)}{q} &\propto
        \ue^{ \frac{\ui}{\hbar} W(q', q; \epsilon) }, \label{eq:quantum_map}
\end{align}
where $W(q', q; \epsilon)$ is the generating function of a one-parameter
family of classical maps acting on a compact phase space $\mathcal{M}^L$.
Thus in the semiclassical limit $\hbar \to 0$ the quantum map reduces to
a classical map $\phi_\epsilon: (q, p) \mapsto (q', p')$ given by the implicit
equations of motion $p = -\partial_q W(q', q; \epsilon)$ and $p' =
\partial_{q'} W(q', q; \epsilon)$, see, e.g.,~\cite{LicLie1992}. The $t$-fold
application of the map is denoted  $x_t \equiv \phi_\epsilon^t x$.
We assume that for $\epsilon = 0$ the corresponding map $\phi_0$ is
non-interacting, i.e., $W(q', q; \epsilon=0) = \sum_{\ell=1}^L
W_\ell^{(0)}(q^{(\ell)}{}^\prime, q^{(\ell)})$. Therefore the subsystem
dynamics splits into
$L$ independent maps $\psi_\ell$, $\ell=1, \ldots, L$, and we take each map to
be chaotic, meaning in our context uniformly hyperbolic.
Without loss of generality, we assume that for $\epsilon > 0$ the
subsystems are coupled. For the moment we leave the specific form of the
interaction and its spatial structure unspecified. Finally, we assume for
simplicity that all discrete symmetries and the time-reversal symmetry of the
system are broken.

\paragraph*{Spectral form factor}
The SFF is a measure of correlations between (quasi)energy levels with
distance $\sim 1/t$. For unitary quantum systems the SFF has the simple form
\begin{align}
 K(t) =  \langle \tr U^t \tr U^{-t} \rangle - \Th^{2} \delta_{t, 0},
 \label{eq:sff_def}
\end{align}
where $\Th = 1/(2\pi\hbar)^L$ is the integer Heisenberg time, which agrees
with the Hilbert space dimension. In the following we omit
the trivial $t=0$ contribution. The average $\langle \cdot \rangle$ is
intrinsically required due to the lack of self-averaging behavior of the SFF
\cite{Pra1997}. It is implicitly realized as, e.g., an average over a
time-window. For semiclassical considerations,
it is natural to rescale time by the mean
level spacing $2 \pi / \Th$, such that $t = \tau \Th$ and $\kappa(\tau) =
K(\tau \Th) / \Th$.
It is a seminal result of periodic orbit theory \cite{Gut1971, HanAlm1984,
Ber1985, SieRic2001, MueHeuBraHaaAlt2004, MueHeuBraHaaAlt2005,
MueHeuAltBraHaa2009}, that the SFF of quantum chaotic single-particle systems
coincides with the random matrix prediction for the corresponding symmetry
classes. For systems with broken time-reversal symmetry, the SFF is
characterized by a linear ramp, $K_\text{RMT}(t) = t$, for times smaller than
the Heisenberg time.

\paragraph*{Semiclassical trace formula} The starting point of a
semiclassical treatment is the evaluation of the path integral for $\tr(U^T)$
via a stationary phase approximation: In the limit $\hbar \to 0$ all fast
oscillations in the path integral are suppressed leaving only stationary
contributions. Those arise from the classical orbits of the collective
dynamics $\phi_\epsilon^t$. This gives rise to Gutzwiller's trace formula
\cite{Gut1971, Tab1983a, HaaGnuKus2018}
\begin{align}
    \tr(U^T) &\sim \sum_{\gamma}
        A_{\gamma} \ue^{\frac{\ui}{\hbar} S_{\gamma}
            - \frac{\ui \pi}{2} \nu_{\gamma} },
        \label{eq:gutzwiller_trace_formula}
\end{align}
where the sum is over all $T$-periodic periodic points $\gamma$
satisfying $\phi_\epsilon^T \gamma = \gamma$ or equivalently over initial
conditions of $T$-periodic orbits $\{ \phi_\epsilon^t \gamma \}_{t=0}^{T-1}$.
The oscillations are generated by the classical action $S_{\gamma} =
\sum_{t=0}^{T-1} W(\gamma_{t+1}, \gamma_{t})$. Each term is weighted by an
amplitude $A_{\gamma} = |\det(M_{\gamma} - 1)|^{-1/2}$ where $M_{\gamma}$ is
the monodromy matrix, characterizing the stability of the orbit. Note that the
periodic points of a hyperbolic map are isolated and unstable.
Lastly, the Maslov indices $\nu_{\gamma}$ account for sign changes of
$\det(M_{\gamma} - 1)$. They cancel in the SFF setting and we omit them in the
following.
The SFF follows directly from the trace formula as a double sum
\begin{align}
 K(T) &\sim \sum_{\gamma, \gamma'} A_{\gamma} A_{\gamma'}
        \ue^{\frac{\ui}{\hbar} ( S_{\gamma} - S_{\gamma'} )}.
        \label{eq:sff_double_sum}
\end{align}
Practically, this double sum cannot be evaluated directly due to the
exponential growing number of unstable $T$-periodic orbits $\sim \ue^{h
T}/T$ as $T \gg 1$, with $h > 0$ the topological entropy \cite{Bow1972,
ParPol1983}. However, assisted by the average in \eqref{eq:sff_def}, rapid
oscillating phase differences $\Delta_{\gamma, \gamma'} = S_{\gamma} -
S_{\gamma'}$ cancel and only pairings with systematically small action
differences $\Delta_{\gamma, \gamma'} \lesssim \hbar$ contribute.
In the following, we discuss shortly the well known and instructive case of
single particle systems: Within the setting of broken time-reversal symmetry,
the only relevant contributions arise from diagonal terms $S_{\gamma} =
S_{\gamma'}$ \cite{Ber1985}.
It is crucial to distinguish between periodic points $\gamma$ and their
corresponding periodic orbits: All iterates $\phi_\epsilon^t \gamma$ of a
periodic point give rise to the same periodic orbit. To account for this
multiplicity we identify the periodic orbit with a distinguished point
$\undga$ arbitrarily chosen from the orbit. Then all periodic points giving
rise to the same periodic orbit are parameterized by $\phi_\epsilon^t
\undga$ for $t \in \mathbb{Z}_T$. In this sense periodic orbits carry a
cyclic symmetry group $\mathbb{Z}_T$ of time translations. The notation is
needlessly complicated here, but will be crucial in the many body setting.
The action $S_\gamma$ and amplitude $A_\gamma$ are invariant under this
symmetry, i.e.,~$S_{\phi_\epsilon^t \undga} = S_{\undga}$ and
$A_{\phi_\epsilon^t \undga} = A_{\undga}$. Thus by making the
multiplicity explicit, one finds $K(T) \sim |\mathbb{Z}_T| \sum_{\gamma}
A_{\gamma}^2$. The remaining sum is known as the Hannay-Ozorio de Almeida sum
rule \cite{HanAlm1984, Par1986, Par1988, PolSha2011}, and yields $\sum_{\gamma}
A_{\gamma}^2 = 1$, thus $K(T) \sim |\mathbb{Z}_T| = T$. In this regard, the SFF
arises by eliminating the physical periodic orbits but keep the degrees of
their symmetry.

\paragraph*{Many-body systems} The implementation of periodic orbit theory for
many-body quantum maps follows from the previous discussion up to
\eqref{eq:sff_double_sum}. However, the parametrically closeness to a
non-interacting configuration has drastic consequences, which we briefly
outline by means of intuitive spectral arguments.
Consider the spectrum of a non-interacting systems as the superposition of the
individual subsystem spectra. Correlations between levels arise only within
levels from the same subsystem spectrum, hence on scales inverse to the
Heisenberg time of the subsystem $T_\text{SH} = \Th^{1/L}$. In this regime the
SFF factorizes as $K(T) = T^L$. On finer scales $\lesssim 1/T_\text{SH}$, the
spectrum is uncorrelated hence Poissonian \cite{TkoSmaKusZeiZyc2012}
corresponding to a plateau $K(T) = \Th$ for $T \geq T_\text{SH}$. By
introducing a small interaction, all levels move and thereby generate
correlations. This process is sensitive
on the scale of the interaction relative to the scale of the spectrum: For
decreasing $\hbar \ll 1$ the mean level distances decreases as well and hence
any finite $\epsilon$ has an increasingly stronger effect on perturbing the
relative distances of the levels. Thus, any finite perturbation is generically
sufficient to induce the standard RMT SFF $K(T) = T$ in the semiclassical
limit.
In contrast, a different behavior arises, when scaling the interaction
accordingly to compensate the decreasing level distances. Perturbation theory
for the random phase model \cite{SriTomLakKetBae2016, LakSriKetBaeTom2016,
TomLakSriBae2018, PulLakSriBaeTom2020, HerKieFriBae2020,
PulLakSriKieBaeTom2023, FriKie2024} predicts that rescaling $\epsilon$ such
that $\sqrt{\Lambda} = \epsilon \sqrt{\Th} / \hbar$ is constant generates
an asymptotically stable regime, where the perturbation of the spectrum acts
homogeneously on all scales of $\hbar$. This differs from standard RMT in that
signatures of the subsystem structure survive for short times, see
Fig.~\ref{fig:periodic_orbits}.

In the following, we show how this intuition from perturbation theory
translates into the setting of periodic orbit theory.
First,
for $\epsilon = 0$ the periodic orbits factorize into $L$ sums over periodic
points $\gamma^{(\ell)}$ of the subsystem dynamics $\psi_\ell$.
Consequentially, we have $K(T; \epsilon = 0) = T^L$.
From the global point of view, the periodic points of the subsystems organize
in families, each represented by a point $\Gamma_0 =
(\underline{\gamma}^{(1)}, \ldots, \underline{\gamma}^{(L)})$. All periodic
points in such a family can be parameterized by $\phi_0^{\vec r} \Gamma_0 =
(\psi_1^{r_1} \underline{\gamma}^{(1)}, \ldots, \psi_L^{r_L}
\underline{\gamma}^{(L)})$, with $\vec r \in \mathbb{Z}_T^L$, see the inset
of Fig.~\ref{fig:periodic_orbits}. Therefore, those
families carry the symmetry $\mathbb{Z}_T^L$ of subsystem time translations.
Furthermore, the uncoupled action is invariant under this symmetry.
Turning to the interacting case, we note that hyperbolicity guarantees
persistence of essentially all periodic orbits under small perturbations
\cite{MeyHalOff2009}, but their locations can vary smoothly. Hence, for $0 <
\epsilon \ll 1$, the families $\Gamma_\epsilon$ are still parameterized by
$\mathbb{Z}_T^L$. The action is no longer invariant under $\mathbb{Z}_T^L$,
but remains invariant under the subgroup of synchronous time-translation
$\phi_\epsilon^{t}$ with $t \in \mathbb{Z}_T$. This indicates
a symmetry breaking from $\mathbb{Z}_T^L$ to $\mathbb{Z}_T$.

For every two distinct periodic orbits $\undga$ and $\undga'$ from the same
family $\Gamma_0$, the action difference is of order $\epsilon$,
i.e,~$\Delta_{\gamma(\epsilon), \gamma'(\epsilon)}(\epsilon) = \epsilon
\Phi_{\gamma(0), \gamma'(0)} + O(\epsilon^2)$ with
\begin{align}
 \Phi_{\gamma, \gamma'}
    &= \sum_{t=0}^{T-1}
            \partial_\epsilon W(\gamma_{t+1}, \gamma_t; \epsilon)
\mid_{\epsilon=0}
    - \partial_\epsilon W(\gamma_{t+1}', \gamma_t'; \epsilon)
\mid_{\epsilon=0},
    \label{eq:approx_action_diff}
\end{align}
see App.~\ref{app:action_diff}. Hence the difference is characterized by the
interaction part of the action in first order of $\epsilon$ evaluated along
the unperturbed orbits in $\Gamma_0$. We argue below, that the action
differences asymptotically depend on $T = \Th \tau$ as $\Phi_{\gamma, \gamma'}
\propto \sqrt{\Th}$ for large periods $T$. Thus by assembling all parametric
dependencies, we recover the universal parameter $\sqrt{\Lambda} = \epsilon
\sqrt{\Th}/ \hbar$ in Eq.~\eqref{eq:sff_double_sum} as predicted by
perturbation theory. Again, when keeping $\epsilon$ finite as $\hbar \to 0$,
only phase differences for $\underline{\gamma} = \underline{\gamma}'$
survive. This suggests the emergence of the standard RMT result.
In contrast, when keeping $\Lambda$ fixed as $\hbar \to 0$, i.e.,~the scaling
limit of vanishing interaction strength, we need to include correlations
between orbits within each family $\Gamma_0$. This limit accurately captures
the interplay of the finite $\hbar$ resolution with $\epsilon$-sized
phase-space structures caused by the interaction. In this regard, this regime
is of particular relevance for characterizing many-body
systems where the numerical value of $\hbar$ is comparable to the
interaction strength. We emphasize that numerical studies \cite{FriKie2024,
ChaLucCha2018a} suggest that the following theory applies, despite its
semiclassical nature, to a wide range of systems at finite $\hbar$.

The effectively vanishing interaction strength $\epsilon \to 0$ allows to
expand all $\epsilon$-dependent terms to first order and keep only terms
which pair with some $\sqrt{\Th} / \hbar$ to the finite $\Lambda$. To this
end, we have
\begin{align}
  K(T) &\sim \sum_{\Gamma_0} A_{\Gamma_0}^2
    \sum_{\vec r, \vec s \in \mathbb{Z}_T^L}
        \exp{\ui \sqrt{\Lambda \tau}
                \tilde{\Phi}_{ \phi_0^{\vec r} \Gamma_0,
                         \phi_0^{\vec s} \Gamma_0 } },
    \label{eq:sff_lambda_limit}
\end{align}
with $\tilde{\Phi}_{ \phi_0^{\vec r} \Gamma_0, \phi_0^{\vec s} \Gamma_0 } =
\Phi_{ \phi_0^{\vec r} \Gamma_0, \phi_0^{\vec s} \Gamma_0
}/\sqrt{\Th \tau}$.
Hence the SFF reduces to a sum over all families $\Gamma_0$ weighted by
some phase factor measuring torsions of orbits within a family. As in the
single particle case, we are aiming for eliminating the periodic
orbits and focus on the degrees of freedom spanned by the
symmetry. This is a conceptionally difficult task as the phases $\tilde{\Phi}$
typically have a
complicated dependency on $\Gamma_0$. However, progress can be made by
invoking the periodic orbit version of a central limit theorem (CLT) for $\Th
\to \infty$. Intuitively, this corresponds to interpreting the phase term as
subject to an average over random variables $\partial_\epsilon W$. Note that
the choice of rescaling the phases by $1/\sqrt{T}$ was the first step in this
direction. Moreover, we prepare \eqref{eq:sff_lambda_limit} for the CLT by
shifting $\vec s \to \vec s + \vec r \mod T$ and reshuffle terms to
\begin{align}
 K(T) &= \sum_{\vec s \in \mathbb{Z}_T^L} R[\tilde{\Phi}_{\vec s}],
 \label{eq:sff_sum_over_s_only_and_functional}
\end{align}
with the functional
\begin{align}
    R[\tilde{\Phi}_{\vec s}] &= \sum_{\vec r} \sum_{\Gamma_0} A_{\Gamma_0}^2
        \exp{\ui \sqrt{\Lambda \tau} \,
                    \tilde{\Phi}_{\vec s}( \phi_0^{\vec r} \Gamma_0) },
        \label{eq:sff_char_fn}
\end{align}
where $\tilde{\Phi}_{\vec s}( \phi_0^{\vec r} \Gamma_0) \equiv
\tilde{\Phi}_{ \phi_0^{\vec r} \Gamma_0, \phi_0^{\vec s + \vec r} \Gamma_0}$.
The transformation of $\vec s$ allows for identifying $\sum_{\vec r}
\sum_{\phi_0^{\vec r} \Gamma_0} = \sum_{\gamma_0}$ as the sum over all
periodic points $\gamma_0$.
Keeping the interpretation as random variables, we can identify
\eqref{eq:sff_char_fn} as the Fourier transform of the probability
distribution of phases $\Phi$. We argue that this probability distribution
tends to a Gaussian distribution as $\Th \to \infty$. In
Ref.~\cite{CanSha2021} a central limit theorem of a trace formula similar to
\eqref{eq:sff_char_fn} but with respect to the maximum entropy measure is
proved. Incorporating the Sinai-Ruelle-Bowen (SRB) measure $\mu(x)$
\cite{Sin1972,Bow1975b,Rue1978} gives
\begin{align}
    R[\tilde{\Phi}_{\vec s}] &\xrightarrow{\Th \to \infty}
        \ue^{-  \sigma_{\Phi_{\vec s}}^2 \Lambda \tau/2}.
    \label{eq:clt_limit}
\end{align}
For a fixed $\vec s$ the variance is given by
\begin{align}
    \sigma_{ \Phi_{\vec s}}^2 &= \lim_{\mathcal{T} \to \infty}
        \frac{1}{\mathcal{T}}
            \int \ud \mu(x) \, \left[ \sum_{t=0}^{\mathcal{T} - 1}
            v_{\vec s}( \phi_0^t x) \right]^2,
    \label{eq:general_variance}
\end{align}
where $v_{\vec s}(x) = \partial_\epsilon W(x) \mid_{\epsilon=0} -
\partial_\epsilon W(\phi_0^{\vec s} x) \mid_{\epsilon=0}$ is the summand of
$\Phi_{\vec s}$. Thus, the CLT transforms the SFF into the partition function
for $\mathbb{Z}_T^L$ degrees of freedom with Hamiltonian $\sigma_{\Phi_{\vec
s}}^2/2$ and inverse temperature $\Lambda \tau$. Such relations to statistical
mechanics problems are ubiquitous in random quantum circuit calculations upon
performing ensemble averages, whereas here they arise from symmetries
acting on periodic orbits.

After eliminating the periodic orbits, the information about the system is
contained in the variance. It can be shown that the variance is constant along
synchronous time translations, i.e.,~shifts $\vec s \to \vec s + t \vec 1 \mod
T$ forming a subgroup $\mathbb{Z}_T$ of $\mathbb{Z}_T^L$, see
App.~\ref{app:symmetry_variance} for details. This is intuitively obvious,
as it is the global time-translation symmetry for which the action in
\eqref{eq:sff_double_sum} is always invariant.
For a quantification of contributions from the broken symmetry degrees of
subsystem time translations $\mathbb{Z}_T^L/\mathbb{Z}_T$, one needs to
specialize the up to here fully general treatment. For concreteness, we
consider the case of a homogeneous nearest-neighbour interaction with
periodic boundary conditions.
In this case, the variance $\sigma_{\Phi_{\vec s}}^2$ remains a sum of
nearest-neighbour terms, see App.~\ref{app:structure_variance}, corresponding
to a local Hamiltonian in the statistical mechanics picture. Additionally, we
assume for simplicity that the interaction function $\partial_{\epsilon} W
\mid_{\epsilon=0}$ has instantaneously decaying classical
correlations, see App.~\ref{app:dyn_variances} for the precise meaning.
This is a strong assumption, which we impose here to have a direct
correspondence to the random matrix model. The more general setting of
(sub)exponentially decaying correlations is discussed in
App.~\ref{app:dyn_variances}.
With this one can show that the effective Hamiltonian is that of the $t$-state
Potts model \cite{Wu1982} $\sigma^2_{\Phi_{\vec s}} = \sigma^2_{\Phi}
\sum_{\ell=1}^{L} (1 - \delta_{s_\ell, s_{\ell+1}})$, where $L + 1 \equiv 1$
and $\sigma^2_{\Phi}$ is some system specific constant, see
App.~\ref{app:structure_variance}.
The same effective model also arises in random quantum-circuit
models with tunable nearest-neighbour interaction  \cite{ChaLucCha2018a}.
Physically, the appearance of the Potts model can be understood in that the
symmetry breaking penalizes asynchronous time translations by an exponential
damping factor $\chi = \ue^{-\Lambda \sigma_\Phi^2/2}$. The detailed
calculation of the SFF from the Potts model is given in
App.~\ref{app:structure_variance} for completeness. In the following, we state
the result for finite, but large $\Th \gg 1$,~where the CLT holds only as an
approximation and discuss the formally correct limit $\Th = 1/(2\pi \hbar)^L
\to \infty$ afterwards. For finite $\Th$, one obtains
\begin{align}
    K(\tau \Th) &= (1 - \chi^\tau + \Th \tau  \chi^\tau)^L +
            (\Th \tau - 1)(1 - \chi^\tau)^L,
            \label{eq:sff_fin_res_unscaled}
\end{align}
which is characterized by a bump-ramp structure, see
Fig.~\ref{fig:periodic_orbits}, identical to the RMT
prediction \cite{ChaLucCha2018a, FriKie2024}. We remark that for systems which
do not have an instantaneous decay of correlations, the overall form of
Eq.~\eqref{eq:sff_fin_res_unscaled} remains. In App.~\ref{app:dyn_variances} we
derive bounds for (sub)exponentially decaying correlations and thereby show
that the appearance of the linear ramp at late times is a generic feature.

It remains to discuss the range of validity of \eqref{eq:sff_fin_res_unscaled}
and a perspective towards the rigorous treatment for $\kappa(\tau) = K(\tau
\Th)/\Th$ as $\Th = 1/(2\pi\hbar)^L \to \infty$. From
\eqref{eq:sff_fin_res_unscaled}, we directly observe that the rescaled SFF
diverges in this limit. At first sight this is attributed to the different
scales $(T_\text{H} \tau)^\ell$ arising in the expansion of the first term of
\eqref{eq:sff_fin_res_unscaled}, but has a more profound origin.
It is due to an ``overcounting'' of long periodic orbits, more precisely of
orbits
which explore the phase space on scales smaller than $\hbar$. Instead, one
needs to implement the hierarchy of self-encounter bunches
\cite{SieRic2001, MueHeuBraHaaAlt2004, MueHeuBraHaaAlt2005} and pseudo-orbits
\cite{MueHeuAltBraHaa2009, WalHeuUrbRic2009, BraHaa2012, MueNov2018a,
MueNov2018b, WalRic2019}: Periodic orbits of parametrically large
period exhibit phase-space structures corresponding to systematically small
action differences. These contributions allow to resolve the plateau, $K(T) =
T_\text{H}$ for $T > T_\text{H}$ in the single particle setting.
Since the orbits entering our computation are effectively elements of the
subsystem phase spaces, such Heisenberg time effects arise already at the
Heisenberg time of the subsystem $T_\text{SH} = T_\text{H}^{1/L}$. By assuming
that the orbits of the subsystems assemble in similar structures, suggests
that after the Heisenberg time of the subsystem, the leading term $\sim T^L$
enters into a plateau. Moreover, the time $T_\text{SH}$, at which the terms
$\sim T^L$ enters into the plateau, becomes asymptotically small in rescaled
variables $\tau_\text{SH} = T_\text{SH}/\Th \to 0$. This suggests
$K(T_\text{SH})/\Th \to \kappa(0) = 1$.
By assuming that a proper resummation establishes the plateau only and
introduces no further contribution, we conjecture $\kappa(\tau) = (1 - y) + y
\tau$ with $y = (1 - \chi^\tau)^{L-1}(1 - \chi^\tau + L \chi^\tau)$ in
correspondence to previous random matrix calculations \cite{FriKie2024,
AltKimMic2025:p}. This interpolates between the expected Poissonian behavior
for the uncoupled case $\Lambda = 0$ ($\chi = 1$) and an effective
single-particle RMT regime for $\Lambda \gg 0$ ($\chi \ll 1$). For
intermediate $\Lambda$ one recovers a linear ramp indicative of random matrix
behavior after the so-called Thouless time $t_\text{TH} \sim
\ln(L)/|\ln(|\chi|)|$ \cite{ChaLucCha2018a}. Thus this extrapolation provides
the full behavior of the SFF.

\paragraph*{Summary and Outlook}
We present a derivation of the SFF for interacting many-body systems with
a semiclassical limit. The SFF identifies a many-body phase that is
characterized by a scaling limit of vanishing interaction strength. The
semiclassical picture builds on the persistence of periodic orbit families and
the emergence of quasi-random fluctuations described by a CLT. These generate a
symmetry breaking of the shift symmetries $\mathbb{Z}_T^L$ to $\mathbb{Z}_T$.
Thereby an effective statistical mechanics model arises, which establishes
a direct correspondence between deterministic systems and random matrix
theory. The SFF coincides with the prediction for random circuits
in the limit of large matrix dimension, i.e.,~small $\hbar$, underlining the
validity of the BGS conjecture in the context of many-body systems.
For a deeper understanding of this quantum chaotic phase it is of central
interest to develop a comprehensive theory of self-encounters and pseudo-orbit
resummations incorporating all subsystem time scales up to the total system
scale.

\paragraph*{Acknowledgements}
We thank J.~D.~Urbina for insightfull discussions. The work has been supported
by Deutsche Forschungsgemeinschaft (DFG), Project No. 497038782 (MK, AB) and
by the European Union's Horizon Europe program under the Marie Sk{\l}odowska
Curie Action GETQuantum (Grant No.~101146632) (FF) and Marie Sk{\l}odowska
Curie Action SemiLiom (Grant No.~101198880) (MK).

\newpage

\appendix

\section{Approximate phase difference} \label{app:action_diff}

In the following we derive the leading term of the action difference given
in Eq.~\eqref{eq:approx_action_diff}. We expand the action to first order in
$\epsilon$. This gives two first order contributions: the first derivative of
the action
evaluated along unperturbed orbits and the non-interacting part of the action
evaluated along the ``derivative of periodic orbits''. In detail, we have for
the action
\begin{align}
    \partial_\epsilon S_{\gamma^{(\epsilon)}}(\epsilon) \mid_{\epsilon=0} &=
    \partial_\epsilon \sum_{t=0}^{T-1}
            W(\gamma_{t+1}^{(\epsilon)}, \gamma_t^{(\epsilon)}; \epsilon)
            \bigg|_{\epsilon=0} \\
    &= s_1 + s_2,
\end{align}
where
\begin{align}
    s_1 &= \sum_{t=0}^{T-1} \frac{\partial W}{\partial \epsilon}
            \bigg|_{\gamma_{t+1}^{(0)}, \gamma_t^{(0)}}
        \equiv (\partial_{\epsilon} S) \mid_{\gamma^{(0)}},
\end{align}
and
\begin{align}
    s_2 &= \sum_{t=0}^{T-1} \left[
        \frac{\partial W(q', q)}{\partial q'}\bigg|_{\gamma_{t+1}, \gamma_t}
                    \partial_\epsilon \gamma_{t+1}
                    \right]_{\epsilon = 0} \\
    &\quad +  \sum_{t=0}^{T-1} \left[
        \frac{\partial W(q', q)}{\partial q}\bigg|_{\gamma_{t+1}, \gamma_t}
                    \partial_\epsilon \gamma_{t} \right]_{\epsilon = 0}.
\end{align}
Using the equations of motion, it follows
\begin{align}
    s_2 &= \sum_{t=0}^{T-1} \left[
            p_{t+1} \partial_\epsilon \gamma_{t+1}
        -   p_{t} \partial_\epsilon \gamma_{t}
                    \right]_{\epsilon = 0}.
\end{align}
By cyclically shifting the first term under the sum from $t+1$ to $t$,
i.e.,~exploiting that $\gamma$ is a periodic orbit, we find $s_2 = 0$ and the
result \eqref{eq:approx_action_diff} in the main text follows.

\section{Symmetry aspect of the variance}
\label{app:symmetry_variance}

We show that the variance defined in Eq.~\eqref{eq:general_variance} is invariant
under global shifts of the from $\vec s \to \vec s + t \vec{1} \mod T$,
i.e.,~$\sigma^2_{\Phi_{\vec s + t \vec 1}} = \sigma^2_{\Phi_{\vec s}}$.
We thereby rely on the following statement \cite{Rob1960, Rat1973, ParPol1990}:
Let
\begin{align}
 \Phi_{\vec s}(x) &= \sum_{t=0}^{T-1} v_{\vec s}(\phi_0^t x),
\end{align}
then $\sigma_{\Phi_{\vec s}}^2 > 0$ if and only if $v_{\vec s}$ is not
cohomologous to a constant $c$, i.e.,~there is no function $g$,
such that $v_{\vec s} + c = g - g \circ \phi_0$. Otherwise the variance is
zero. We call $h = g - g \circ \phi_0$ a coboundary. The intuition behind
this is, that a coboundary gives a constant (more correctly an $O(1)$)
contribution in the time summation in \eqref{eq:general_variance}. Due to the
overall weight $1/\mathcal{T}$, this constant contribution vanishes in the
limit $\mathcal{T} \to \infty$. Similarly we have that adding a coboundary to
$v_{\vec s}$ does not alter the variance. From this it is straightforward to
show the claim as
\begin{align}
 v_{\vec s + t \vec 1} = v_{\vec s} + g - g \circ \phi_0,
\end{align}
with
\begin{align}
 g(x) = \sum_{n=0}^{t - 1} (\partial_\epsilon W)(\phi_0^{\vec s + n \vec 1} x).
\end{align}
Therefore, we can restrict the $\vec s$ dependence to (equivalence classes of)
shifts from the quotient group $[\vec s] \in \mathbb{Z}_T^L / \mathbb{Z}_T
\simeq \mathbb{Z}_T^{L-1}$. To be more explicit, one might consider the
mapping $\pi: \mathbb{Z}_T^L \to \mathbb{Z}_T^{L-1}$ given by $(s_1, \ldots,
s_L) \mapsto (s_2 - s_1, \ldots s_L - s_1)$. It is a group homomorphism and
above group quotient follows from the isomorphism theorem by $\mathbb{Z}_T^L /
\ker(\pi) \simeq \text{im}(\pi)$. Thus one can denote a point in
$\mathbb{Z}_T^{L-1}$ as an image of some point $\vec s \in \mathbb{Z}_T^{L}$
under $\pi$, i.e., $[\vec s] \equiv \pi(\vec s)$. Intuitively, this means that
we gauge out from every $\vec s$ the part which is $\propto \vec 1$. Every
equivalence class $[\vec s] \in \mathbb{Z}_T^{L-1}$ has $T$ members, but by
invariance of the variance, each member gives the same contribution and we have
\begin{align}
 K(T) = T \sum_{[\vec s] \in \mathbb{Z}_T^{L-1}} \ue^{-\Lambda \tau
\sigma^2([\vec s])/2}.
\end{align}
Note that we do not use this notation in the main text, but rely on it in the
following appendices for convenience.

Further structural features of the interaction as, e.g., a nearest-neighbour
coupling can constrain the sum over shifts even further. However, this does not
cleanly affect the group theoretic aspect and hence is separately considered in
App.~\ref{app:structure_variance}.

\section{Structural aspect of the variance}
\label{app:structure_variance}

We show that for a nearest-neighbour interaction of the form
\begin{align}
    \partial_\epsilon W \mid_{\epsilon = 0} &= \sum_{\ell=1}^L
        w(x_\ell, x_{\ell + 1}) \qquad L+1 \equiv 1,
\end{align}
combined with instantaneously decaying classical correlations, the variance
corresponds to the Hamiltonian of the $t$-state Potts model
\begin{align}
 \sigma_{\Phi_{\vec s}}^2 &= \sigma_{\Phi}^2 \sum_{\ell=1}^{L}
            (1 - \delta_{s_\ell, s_{\ell + 1}}).
\end{align}
Furthermore, we present a short derivation for expressing the partition
function of the Potts model in a closed form.

The proof follows in two steps. First we show that the variance preserves the
nearest-neighbour interaction of the physical system and thus the effective
Hamiltonian remains nearest-neighbour interacting. Afterwards the Potts model
is solved for this Hamiltonian via a transfer matrix calculation. We set
\begin{align}
    v_{\vec s}(x) = \sum_{\ell = 1}^L v_{s_\ell, s_{\ell + 1}}(x_\ell,
x_{\ell + 1}),
\end{align}
with
\begin{align}
 v_{s_\ell, s_{\ell + 1}}(x_\ell, x_{\ell + 1}) &=
w(x_\ell, x_{\ell + 1}) - w( \psi_{\ell}^{s_{\ell}} x_\ell,
\psi_{\ell+1}^{s_{\ell + 1}} x_{\ell + 1}).
\end{align}
We start  by inserting the nearest-neighbour interaction into
\eqref{eq:general_variance} and interchange the integral and summation over
subsystems to obtain
\begin{align}
 \sigma_{\Phi_{\vec s}}^2 &= \lim_{\mathcal{T} \to \infty}
        \frac{1}{\mathcal{T}} \sum_{\ell, \ell' = 1}^L \int \ud \mu(x)
            K_\ell(x, \mathcal{T})
            K_{\ell'}(x, \mathcal{T}),
    \label{eq:app_variance_with_nn_inter}
\end{align}
with
\begin{align}
     K_\ell(x, \mathcal{T}) &= \sum_{t=0}^{\mathcal{T} - 1}
            v_{s_\ell, s_{\ell+1}}(\psi_\ell^t x_\ell,
            \psi_{\ell+1}^t x_{\ell+1}).
\end{align}
Note that by invariance of the measure we have
\begin{align}
    \int \ud \mu(x) K_\ell(x, \mathcal{T}) &= 0.
\end{align}
Hence all terms of the double sum in \eqref{eq:app_variance_with_nn_inter} for
which the integrals factorize vanish. Only three terms $\ell' \in \{ \ell - 1,
\ell, \ell + 1 \}$ survive. In the following, we argue that also the terms
$\ell' = \ell \pm 1$ vanish. Consider first the case $\ell' = \ell + 1$,
i.e.,~terms of the form
\begin{align}
    (\sigma_{\Phi_{\vec s}}^2)^+ &= \lim_{\mathcal{T} \to \infty}
        \frac{\mathcal{V}^{L-3}}{\mathcal{T}} \sum_{\ell=1}^L
            \int \ud \mu(x_{\ell}, x_{\ell+1}, x_{\ell+2}) I_\ell,
\end{align}
where $\mathcal{V}$ is the phase-space volume of a subsystem (we assumed
compact
phase spaces, hence the phase-space volume is finite and implicitly assumed to
be normalized to unity) and
\begin{align}
    I_\ell &= \overline{v_{s_\ell, s_{\ell+1}}}^{\mathcal{T}} \cdot
              \overline{v_{s_{\ell+1}, s_{\ell+2}}}^{\mathcal{T}},
\end{align}
where we omitted the arguments to keep notation simple and indicated
the sum over times by the overline. Clearly, $x_\ell$ occurs only in the first
term and $x_{\ell+2}$ occurs exclusively in the second term. We define
\begin{align}
 w^+(y) &= \int \ud \mu(x) w(x, y) \\
 w^-(x) &= \int \ud \mu(y) w(x, y),
\end{align}
and find
\begin{align}
  (\sigma_{\Phi_{\vec s}}^2)^+ &= \lim_{\mathcal{T} \to \infty}
\frac{\mathcal{V}^{L-3}}{\mathcal{T}} \sum_{\ell=1}^L
            \int \ud \mu(x_{\ell+1}) \iota_+(x_{\ell+1}) \iota_-(x_{\ell+1}),
\end{align}
with
\begin{align}
 \iota_{\pm}(x) &= \sum_{t=0}^{\mathcal{T} - 1} \left(
        w^{\pm}(\psi_{\ell+1}^t x) - w^{\pm}(\psi_{\ell+1}^{t+s_{\ell+1}} x)
\right).
\end{align}
Here, $g^\pm(x) = w^{\pm}(x) - w^{\pm}(\psi_{\ell+1}^{s_{\ell+1}} x)$ are
coboundaries, more precisely, $g^\pm(x) = h^{\pm}(x) - h^\pm(\psi x)$ with
\begin{align}
 h^\pm(x) &= \sum_{t=0}^{s_{\ell+1}-1} w^{\pm}(\psi_{\ell+1}^t x).
\end{align}
As already explained in App.~\ref{app:symmetry_variance}, the variance vanishes
on coboundaries, as
\begin{align}
    \sum_{t=0}^{\mathcal{T} - 1} (h^\pm(\psi^t x)
        - h^\pm(\psi^{t+1} x)) &= h^\pm(x) - h^\pm(\psi^\mathcal{T} x),
\end{align}
hence the summation over time reduces to a finite summation over $s$ only.
Note that it is crucial, to take the limit $\mathcal{T} \to \infty$
independently from $\Th \to \infty$. By invoking for instance a mixing
property, one observes that the integral remains finite and due to the
weighting by $1/\mathcal{T}$ the term vanishes. The argumentation holds
analogously for $\ell' = \ell - 1$. Only the term $\ell' = \ell$ survives and
hence the variance decomposes as
\begin{align}
 \sigma_{\Phi_{\vec s}}^2 &= \sum_{\ell=1}^L \sigma^2_{v_{s_\ell, s_\ell+1}}.
\end{align}
The above shows, that in the statistical mechanics picture the SFF corresponds
to a local Hamiltonian with essentially arbitrary nearest-neighbour
interactions.

The nearest-neighbour interaction allows to establish a transfer matrix
approach to the SFF. The exponential involving the variance factorizes, i.e.,
\begin{align}
    \exp{-\Lambda \tau \sigma^2_{\Phi_{\vec s}}/2} =
            \prod_{\ell = 1}^L \exp{- \frac{\Lambda \tau
                \sigma^2_{v_{s_\ell, s_{\ell+1}}}}{2}},
\end{align}
and hence it can be written as a trace $K(T) = \tr \Omega_T^L$ over transfer
matrices
\begin{align}
    (\Omega_T)_{s_\ell, s_{\ell'}} = \exp{-\frac{\Lambda \tau
                \sigma^2_{v_{s_\ell, s_{\ell+1}}}}{2}}.
\end{align}
Furthermore, it can be shown that the variances $\sigma^2_{v_{s_\ell,
s_{\ell+1}}}$ depend up to a vanishing coboundary on relative shifts
$\tilde{s}_\ell = s_\ell - s_{\ell + 1} \mod T$ only. Hence the matrix depends
only on the relative shift $\tilde{s} = s_\ell - s_{\ell + 1}$ as well. The
transfer operator thus is a circulant matrix and can be diagonalized by the
discrete Fourier transform. The eigenvalues $\{ \lambda_n
\}_{n=0}^{T-1}$ of circulant matrices are generally expressible as
\begin{align}
    \lambda_n &= \sum_{\tilde{s} \in \mathbb{Z}_T}
        \exp{-\frac{\Lambda \tau \sigma^2_{v_{\tilde{s}}}}{2}
                + \frac{2 \pi \ui n \tilde{s}}{T} }.
\end{align}
Given the eigenvalues the SFF follows by $K(T) = \sum_{n} \lambda_n^L$.
Note that this applies to an arbitrary nearest-neighbour interaction. However,
the dependence of $\sigma^2_{v_{\tilde{s}}}$ on $\tilde{s}$ can be complicated
hence preventing progress towards a closed form solution for the eigenvalues.

Therefore, we restrict in the main text to the case where the resulting
statistical mechanics model is the exactly solvable Potts model. In
App.~\ref{app:dyn_variances}, we discuss in more detail, that the variances
typically decay for $|\tilde{s}| > 0$. Instantaneous decay of correlations
refers to $\sigma^2_{v_{s_\ell, s_{\ell+1}}} = \sigma_{v}^2 (1 -
\delta_{s_\ell - s_{\ell + 1}, 0})$, i.e.,~already at $|\tilde{s}| = 1$ the
$\vec s$-dependent contributions have decayed completely. By denoting $\chi
= \ue^{-\Lambda \sigma^2_{v}/2}$ the eigenvalues calculate to $\lambda_0
= 1 - \chi^\tau + T \chi^\tau$ and $\lambda_n = 1 - \chi^\tau$ for the
remaining eigenvalues with $n = 1, \ldots, T-1$. Hence, the SFF reduces as
\begin{align}
    K(T) &= \sum_{n = 0}^{T - 1} (\lambda_n)^L \\
         &= (1 - \chi^\tau + T \chi^\tau)^L + (T - 1) (1 - \chi^\tau)^L,
\end{align}
which is the result \eqref{eq:sff_fin_res_unscaled} stated in the main text.

\section{Dynamical aspect of the variance}
\label{app:dyn_variances}

We discuss here the dynamical features entering the variance in more detail.
Therefore, we give first a convenient relation of the variance to a function
involving classical correlation functions only. This is particularly important
for relating the decay of asynchronous shifts to the decay of classical
correlations. We present an argumentation in this direction by calculating
the corrections from subexponentially decaying correlations to the SFF for
instantaneously decaying correlations. Those calculations are carried out for
simplicity in the setting of an all-to-all interacting system with open
boundaries. With minor adaptions, the below arguments equally well apply to
nearest-neighbour interactions.

The relation for the variance can be stated completely general, without
reference to any structure of the interaction. Therefore, let $v_{\vec s} = w
- w \circ \phi_0^{\vec s}$ for some function $w$ for which we assume without
loss of generality $\langle w(x) \rangle_\mu = 0$, where $\langle \cdot
\rangle_\mu$ is the phase-space average w.r.t.~the SRB measure. Otherwise, we
set $w \to w - \langle w \rangle_\mu$. The variance can be written as
\cite{Leo1961, Rat1973}
\begin{align}
     \sigma_{v_{\vec s}}^2 &= 2 \sum_{t=-\infty}^{\infty}
            \left( C_w(t \vec 1 ) - C_w( t \vec 1 + \vec s) \right),
     \label{eq:dynamical_variance}
\end{align}
where $C_w(\vec s) = \langle w(\phi_0^{\vec s} x) w(x) \rangle_\mu$ is the
correlation function of a function $w$ w.r.t.~the SRB measure.
The expression \eqref{eq:dynamical_variance} holds if $C_{w - w \circ
\phi_0^{\vec s}}(t)$ decays at least subexponentially, i.e.,~$|C_{w - w \circ
\phi_0^{\vec s}}(t)| \leq c \eta^{t^\theta}$ for some $0 < \eta < 1$ and
$\theta > 0$.

In the main text we assumed an instantaneous decay of correlations. This is a
strong requirement, which guarantees that the phase differences behave as
independent random variables. In contrast, for real physical systems a
(sub)exponential decay is more common. Therefore, we derive in the following a
bound on the deviation of the SFF for subexponentially decaying correlations
from the SFF for instantaneous decay of correlations.

It is straightforward to show that for an all-to-all interaction, the SFF
attains the form
\begin{align}
 K(T) &\leq T + T
    \sum_{ [\vec s] \in \mathbb{Z}_T^{L-1} \setminus \{ [\vec 0] \}}
        \exp{-\frac{\Lambda \tau}{2} \sigma_{v_{[\vec s]}}^2 }.
\end{align}
Furthermore, instantaneous decay of correlations refers here to
\begin{align}
    C_w(\vec s) &= \begin{cases}
                         C_w(\vec s) & \vec s = s \vec 1 \\
                         0 & \text{otherwise}
                   \end{cases},
\end{align}
i.e., the correlation function vanishes instantaneously for asynchronous
shifts, but can be of slower decay along synchronous time translations. In
consequence the variance is given by
\begin{align}
    (\sigma^{(0)}_{v_{\vec s}})^2 &= \begin{cases}
                                0 & \vec s = s \vec 1 \\
                                f([\vec 0]) & \text{otherwise}
                             \end{cases},
\end{align}
where we define
\begin{align}
 f([\vec s]) &= 2 \sum_{t=-\infty}^{\infty} C_w(t \vec 1 + \vec s).
\end{align}
The SFF for instantaneously decaying correlations (corresponding again to the
RMT prediction \cite{FriKie2024}) is given by
\begin{align}
 K_0(T) &= T + ( T^L - T) \exp{-\frac{\Lambda \tau f([\vec 0])}{2}}.
\end{align}
In particular, for late times $T \gg 1$ the RMT SFF $K_0(T) = T$ is
recovered.

In the remainder, we show that the RMT regime of the SFF arises also for
subexponentially decaying correlations, i.e.,~is a generic feature of the SFF
for the considered many-body systems. Therefore, we estimate the deviation $|
K(T) - K_0(T)|$ under two assumptions: Firstly, the correlations are bounded
by a subexponential decay, i.e.,~$|C_w(\vec s)| \leq c \eta^{|\vec
s|^{\theta}}$ for some $c, \theta > 0$ and $0 < \eta < 1$. Here we use
the $1$-norm, $|\vec s| = \sum_{\ell=1}^L |s_\ell|$. Secondly, we assume that
$v_{\vec s}$ for $\vec s \in \mathbb{Z}_T^L$ is cohomologous to a constant if
and only if $\vec s \propto \vec 1$. From this it is proved below, that the
variance is bounded by two constants $a, A > 0$ such that
\begin{align}
    a < \sigma_{v_{\vec s}}^2  < A \label{eq:bound_sigma} \\
    a < (\sigma_{v_{\vec s}}^{(0)})^2  < A, \label{eq:bound_sigma0}
\end{align}
for $\vec s \not \propto \vec 1$.

Before deriving these bounds we first estimate the deviation of the SFF by
\begin{align}
 |K(T) - K_0(T)| &= \left| \sum_{\vec s \in \mathbb{Z}_T^L}
    \ue^{-\frac{\Lambda\tau}{2} \sigma_{v_{\vec s}}^2}
    - \ue^{-\frac{\Lambda\tau}{2} (\sigma_{v_{\vec s}}^{(0)})^2} \right| \\
    &\leq \sum_{\vec s \in \mathbb{Z}_T^L} \left|
    \ue^{-\frac{\Lambda\tau}{2} \sigma_{v_{\vec s}}^2}
    - \ue^{-\frac{\Lambda\tau}{2} (\sigma_{v_{\vec s}}^{(0)})^2} \right|.
\end{align}
Note that terms with $\vec s \propto \vec 1$ cancel. The term under the sum
can be estimated using the intermediate value theorem such that
\begin{align}
 \left| \ue^{-\frac{\Lambda\tau}{2} \sigma_{v_{\vec s}}^2}
    - \ue^{-\frac{\Lambda\tau}{2} (\sigma_{v_{\vec s}}^{(0)})^2} \right|
    &= \frac{\Lambda \tau}{2} \ue^{-\frac{\lambda \tau x}{2}}
        \left| \sigma_{v_{\vec s}}^2 - (\sigma_{v_{\vec s}}^{(0)})^2
\right|,
\end{align}
for some $x$ between $\sigma_{v_{\vec s}}^2 = f([\vec 0]) - f([\vec s])$ and
$(\sigma_{v_{\vec s}}^{(0)})^2 = f([\vec 0])$. As we have $a <
\sigma_{v_{\vec s}}^2$ and $a < (\sigma_{v_{\vec s}}^{(0)})^2$ for $\vec
s \not \propto \vec 1$, and the exponential is monotonously decreasing, the
term can be bounded by choosing $x = a$, hence
\begin{align}
 \left| \ue^{-\frac{\Lambda\tau}{2} \sigma_{v_{\vec s}}^2}
    - \ue^{-\frac{\Lambda\tau}{2} (\sigma_{v_{\vec s}}^{(0)})^2} \right|
    &< \frac{\Lambda \tau}{2} \ue^{-\frac{\lambda \tau a}{2}}
        \left| \sigma_{v_{\vec s}}^2 - (\sigma_{v_{\vec s}}^{(0)})^2
      \right| \\
    &< \frac{\Lambda \tau A}{2} \ue^{-\frac{\lambda \tau a}{2}}.
\end{align}
By evaluating the sum over $\mathbb{Z}_T^L$ one finds
\begin{align}
  |K(T) - K_0(T)| &< (T^L - T)
        \frac{\Lambda \tau A}{2} \ue^{-\frac{\lambda \tau a}{2}}.
\end{align}
The deviations are exponentially suppressed for large times and vanish for $T
\to \infty$. Therefore, the linear ramp, indicative for a RMT SFF, survives
even for the case of subexponentially decaying correlations.

It remains to justify the bounds \eqref{eq:bound_sigma} and
\eqref{eq:bound_sigma0}. We note that the constant $f([\vec 0])$ follows from
the variance \eqref{eq:general_variance}, by choosing $w$,
instead of $v_{\vec s} = w - w \circ \phi_0^{\vec s}$. Therefore, by the same
arguments of \cite{Leo1961, Rat1973}, we have $f([\vec 0]) \geq 0$, with
equality if and only if $w$ is cohomologous to a constant. We show that
$f([\vec 0]) > 0$ follows directly from the more general assumption that
$v_{\vec s} = w - w \circ \phi_0^{\vec s}$ is cohomologous to a constant if and
only if $\vec s \propto \vec 1$: For the sake of contradiction, we assume $w$
is cohomologous to a constant $d$, i.e.,~$w = d + g - g \circ \phi_0$, then
for an arbitrary choice $\vec s \not\propto \vec 1$
\begin{align}
    v_{\vec s} &= w - w \circ \phi_0^{\vec s} \\
      &= g - g \circ \phi_0
        - g \circ \phi_0^{\vec s} + g \circ \phi_0^{\vec s + \vec 1}.
\end{align}
By defining $h_{\vec s} = g - g \circ \phi_0^{\vec s}$ it follows
\begin{align}
 v_{\vec s} &= h_{\vec s} - h_{\vec s} \circ \phi_0.
\end{align}
Thus $v_{\vec s}$ is for $\vec s \not\propto \vec 1$ cohomologous to $0$,
which contradicts the premise. Hence, we conclude $f([\vec 0]) > 0$.

From the subexponential decay of correlations we have
\begin{align}
 |f([\vec s])| &= 2 \left| \sum_{t=-\infty}^{\infty}
                    C_w(t \vec 1 + \vec s) \right|
        < c' \eta^{|\vec s|^\theta_\perp},
\end{align}
where $|\vec s|_\perp$ indicates the minimal distance of $\vec s$ to the
diagonal $\mathbb{Z}_T$ in $\mathbb{Z}_T^L$. This estimate gives an upper
bound on $\sigma_{v_{\vec s}}^2 = f([\vec 0]) - f([\vec s])$ as
\begin{align}
    \sigma_{v_{\vec s}}^2 \leq f([\vec 0]) + |f([\vec s])|
                            < f([\vec 0]) + c' =: A.
\end{align}
It remains to justify the lower bound. We observe that the largest
contributions come from $\vec s$ close to the diagonal $\propto \vec 1$.
Therefore, we fix a $r$ such that for all $\vec s$ with $|\vec s|_\perp > r$,
one
has $f([\vec s]) < \frac{f([\vec 0])}{2}$ and hence
\begin{align}
    \sigma_{v_{\vec s}}^2 > \frac{f([\vec 0])}{2}.
\end{align}
Furthermore by assumption that $v_{\vec s}$ is cohomologous to a constant if
and only if $\vec s \propto \vec 1$ it follows that $\sigma_{v_{\vec s}}^2 >
0$ for $\vec s \not \propto \vec 1$. In addition, there are only finitely many
values $[\vec s]$ for which $|\vec s|_\perp \leq r$. Note, that the above
means that we need to impose only finitely many assumptions on $v_{\vec s}$ as
the decay of correlations implies that the latter is not cohomologous to a
constant for almost all $\vec s \in \mathbb{Z}^L$. The above
considerations guarantee that there exists a minimum such that
\begin{align}
    \min_{0 < |\vec s|_\perp \leq r} \sigma_{v_{\vec s}}^2 =: \tilde{a} > 0.
\end{align}
By setting $a = \min \{ \tilde{a}, f([\vec 0]) / 2 \}$, the bound
\eqref{eq:bound_sigma} follows. It is straightforward to check from the
explicit definition of $a$ and $A$ that \eqref{eq:bound_sigma0} holds as
well. This concludes the proof.


\end{document}